\documentclass[11pt]{article}
\usepackage{graphicx,amsfonts,amsmath,amssymb,hyperref}
\begin{document}

\title{Macro- and Microscopic Self-Similarity in Neuro- and Psycho-Dynamics}
\author{{Vladimir G. Ivancevic}\\
{\small Defence Science \& Technology Organisation, Australia}\\
{Tijana T. Ivancevic}\\
{\small School of Electrical and Information Engineering,
University of South Australia, Australia}}\date{}\maketitle

\begin{abstract}
The unique Hamiltonian description of neuro- and psycho-dynamics
at the macroscopic, classical, inter-neuronal level of brain's
neural networks, and microscopic, quantum, intra-neuronal level of
brain's microtubules, is presented in the form of open Liouville
equations. This implies the arrow of time in both neuro- and
psycho-dynamic processes and shows the existence of the formal
neuro-biological space-time self-similarity.\newline

Hamiltonian and Liouville dynamics,
Neuro-biological self-similarity
\end{abstract}

\section{Introduction}

Neuro-- and psycho--dynamics have its physical behavior both on the
{macroscopic}, classical, {inter--neuronal} level
\cite{VladNick,CoMind}, and on the {microscopic}, quantum,
{intra--neuronal} level \cite{LifeSpace,NeuFuz,QuLeap}. On the
{macroscopic} level, various models of neural networks (NNs, for
short) have been proposed as goal--oriented models of the specific
neural functions, like for instance, function--approximation,
pattern--recognition, classification, or control (see, for
example~\cite{A}). In the physically--based, Hopfield--type models
of NNs \cite{B1,B2} the information is stored as a
content--addressable
memory in which synaptic strengths are modified after the Hebbian rule (see~%
\cite{C}). Its retrieval is made when the network with the
symmetric couplings works as the point--attractor with the fixed
points. Analysis of both {activation} and {learning dynamics} of
Hopfield--Hebbian NNs using the techniques of statistical
mechanics~\cite{D}, provides us with the most important
information of storage capacity, role of noise and recall
performance \cite{NatBiodyn,CoMind}.

Conversely, an indispensable role of quantum theory in the brain
dynamics was emphasized in \cite{Kurita}. On the general
{microscopic} intra--cellular level, energy transfer across the
cells, without dissipation, had been first conjectured to occur in
biological matter by~\cite{E}. The phenomenon conjectured by them
was based on their 1D superconductivity model: in one
dimensional electron systems with holes, the formation of
{solitonic structures} due to electron--hole pairing results in the
transfer of electric current without dissipation. In a similar manner, Fr%
\"{o}lich and Kremer conjectured that energy in biological matter
could be transferred without dissipation, if appropriate solitonic
structures are formed inside the cells. This idea has lead
theorists to construct various models for the energy transfer
across the cell, based on the formation of {kink} classical
solutions \cite{F1,F2}.

The interior of living cells is structurally and dynamically
organized by {cytoskeletons}, i.e., networks of protein polymers.
Of these structures, {microtubules} (MTs, for short) appear to be
the most fundamental \cite{G}. Their dynamics has been studied by
a number of authors in connection with the mechanism responsible
for dissipation-free energy transfer. Hameroff and his colleagues
\cite{H1,H2,H3,H4} have conjectured another fundamental role for
the MTs, namely being responsible for {quantum computations} in
the human neurons. Penrose \cite{I1,I2,I3,I4} further argued that
the latter is associated with certain aspects of quantum theory
that are believed to occur in the cytoskeleton MTs, in particular
quantum superposition and subsequent collapse of the wave function
of coherent MT networks. These ideas have been elaborated
by~\cite{J} and \cite{K}, based on the quantum--gravity language
of~\cite{L}, where MTs have been physically modelled as
non--critical (SUSY) bosonic strings. It has been suggested that
the neural MTs are the microsites for the emergence of stable,
macroscopic quantum coherent states, identifiable with the
{preconscious states}; stringy--quantum space-time effects trigger
an organized collapse of the coherent states down to a specific or
{conscious state}. More recently, the evidence for biological
self-organization and pattern formation during embryogenesis was
presented in \cite{M}.

In particular, MTs in the cytoskeletons of eukaryotic cells
provide a wide range of micro--skeletal and micro--muscular
functionalities. Some evidence has indicated that they can serve
as a medium for intracellular signaling processing. For the
inherent symmetry structures and the electric properties of
tubulin dimers, the microtubule (MT) was treated as a
1D ferroelectric system in \cite{Chen}. The nonlinear
dynamics of the dimer electric dipoles was described by virtue of
the double--well potential and the physical problem was further
mapped onto the pseudo--spin system, taking into account the effect
of the external electric field on the MT.

More precisely, MTs are polymers of tubulin subunits (dimers)
arranged on a hexagonal lattice. Each tubulin dimer comprises two
monomers, the $\alpha-$tubulin and $\beta-$tubulin, and can be found in two
states. In the first state a mobile negative charge is located
into the $\alpha-$tubulin monomer and in the second into the $\beta-$tubulin
monomer. Each tubulin dimer is modelled as an electrical dipole
coupled to its neighbors by electrostatic forces. The location of
the mobile charge in each dimer depends on the location of the
charges in the dimer's neighborhood. Mechanical forces that act on
the microtubule affect the distances between the dimers and alter
the electrostatic potential. Changes in this potential affect the
mobile negative charge location in each dimer and the charge
distribution in the microtubule. The net effect is that mechanical
forces affect the charge distribution in microtubules
\cite{Karafyllidis}.

Various models of the mind have been based on the idea that neuron
MTs can perform computation. From this point of view, information
processing is the fundamental issue for understanding the brain
mechanisms that produce consciousness. The cytoskeleton polymers
could store and process information through their dynamic coupling
mediated by mechanical energy. The problem of information transfer
and storage in brain microtubules was analyzed in \cite{Faber},
considering them as a communication channel.

Therefore, we have two space-time biophysical scales of neuro- and
psycho-dynamics: classical and quantum. Naturally the question
arises: are these two scales somehow inter-related, is there a
space-time self-similarity between them?

The purpose of the present paper is to prove the formal positive
answer to the self-similarity question. We try to describe
neurodynamics on both physical levels by the {unique form} of a
single equation, namely {open Liouville equation}: NN--dynamics
using its classical form, and MT--dynamics using its quantum form
in the Heisenberg picture. If this formulation is consistent, that
would prove the {existence} of the {formal neuro-biological
space-time self-similarity}.

\section{Mathematics of Open Liouville Equation}

\subsection{Hamiltonian framework}

Suppose that on the macroscopic NN--level we have a conservative Hamiltonian
system acting in a $2N$D symplectic phase space $T^{\ast
}Q=\{q^{i}(t),p_{i}(t)\},\,i=1\dots N$ \ (which is the cotangent bundle of
the NN--configuration manifold $Q=\{q^{i}\}$), with a Hamiltonian function $%
H=H(q^{i},p_{i},t):T^{\ast }Q\times \mathbb{R}\rightarrow
\mathbb{R}$ (see \cite{VladSIAM,HumLike,GeoDyn}). The conservative
dynamics is defined by classical Hamilton's canonical equations :
\begin{align}
\dot{q}^{i}& =\partial _{p_{i}}H\quad \text{-- contravariant velocity
equation}\,,  \notag \\
\dot{p}_{i}& =-\partial _{q^{i}}H\quad \text{-- covariant force equation}\,%
\text{,}  \label{Ham}
\end{align}%
(here and henceforth overdot denotes the total time derivative). Within the
framework of the conservative Hamiltonian system (\ref{Ham}) we can apply
the formalism of classical Poisson brackets: for any two functions $%
A=A(q^{i},p_{i},t)$ and $B=B(q^{i},p_{i},t)$ their Poisson bracket
is (using the summation convention) defined as \cite{GeoDyn,GaneshADG}
\begin{equation*}
\lbrack A,B]=(\partial _{q^{i}}A\,\partial _{p_{i}}B-\partial
_{p_{i}}A\,\partial _{q^{i}}B).
\end{equation*}

\subsection{Conservative classical system}

Any function $A(q^{i},p_{i},t)$ is called a {constant} (or
integral) of motion of the conservative system (\ref{Ham}) if
\cite{GeoDyn,GaneshADG}
\begin{equation}
\dot{A}\equiv \partial _{t}A+[A,H]=0,\qquad \text{which implies}\qquad
\partial _{t}A=-[A,H]\,.  \label{conserv}
\end{equation}%
For example, if $A=\rho (q^{i},p_{i},t)$ is a {density function of
ensemble phase--points} (or, a probability density to see a state $\mathbf{x}%
(t)=(q^{i}(t),p_{i}(t))$ of {ensemble} at a moment $t$), then
equation
\begin{equation}
\partial _{t}\rho =-[\rho ,H]  \label{LioCl}
\end{equation}%
represents the \textit{Liouville theorem}, which is usually derived from
the {continuity equation}
\begin{equation*}
\partial _{t}\rho +div(\rho \,\mathbf{\dot{x}})=0\,.
\end{equation*}

\subsection{Conservative quantum system}

We perform the formal quantization of the conservative equation (\ref{LioCl}%
) in the Heisenberg picture: all variables become Hermitian operators
(denoted by `$\wedge $'), the symplectic phase space $T^{\ast
}Q=\{q^{i},p_{i}\}$ becomes the Hilbert state space $\mathcal{H}=\mathcal{H}%
_{\hat{q}^{i}}\otimes \mathcal{H}_{\hat{p}_{i}}$ (where $\mathcal{H}_{\hat{q}%
^{i}}=\mathcal{H}_{\hat{q}^{1}}\otimes ...\otimes
\mathcal{H}_{\hat{q}^{N}}$ and
$\mathcal{H}_{\hat{p}_{i}}=\mathcal{H}_{\hat{p}_{1}}\otimes
...\otimes \mathcal{H}_{\hat{p}_{N}}$), the classical Poisson
bracket $[\,,\,]$ becomes the quantum commutator $\{\,,\,\}$
multiplied by $-i/\hbar $ \cite{ComDyn,QuLeap}
\begin{equation}
\lbrack \,,\,]\longrightarrow -i\{\,,\,\}\,\qquad (\hbar =1\,\,\text{in
normal units})\,.  \label{qu}
\end{equation}%
In this way the classical Liouville equation (\ref{LioCl}) becomes
the {quantum Liouville equation} \cite{GaneshADG,QuLeap}
\begin{equation}
\partial _{t}\hat{\rho}=i\{\hat{\rho},\hat{H}\}\,,  \label{LioQu}
\end{equation}%
where $\hat{H}=\hat{H}(\hat{q}^{i},\hat{p}_{i},t)$ is the Hamiltonian
evolution operator, while
\begin{equation*}
\hat{\rho}=\sum_{a}P(a)|\Psi _{a}><\Psi _{a}|,\quad \text{with}\quad \mathrm{%
Tr}(\hat{\rho})=1
\end{equation*}%
denotes the von Neumann {density matrix operator}, where each
quantum
state $|\Psi _{a}>$ occurs with probability $P(a)$; $\hat{\rho}=\hat{\rho}(%
\hat{q}^{i},\hat{p}_{i},t)$ is closely related to another von
Neumann concept: {entropy} $$S=-\mathrm{Tr}(\hat{\rho}[\ln
\hat{\rho}]).$$

\subsection{Open classical system}

We now move to the {open} (nonconservative) system: on the
macroscopic NN--level the {opening operation} equals to the {%
adding} of a {covariant} vector of external (dissipative and/or
motor) forces $F_{i}=F_{i}(q^{i},p_{i},t)$ to (the right-hand-side
of) the covariant Hamilton's {force equation}, so that Hamilton's
equations obtain the {open} (dissipative and/or forced) form
\cite{VladSIAM,HumLike,GeoDyn}:
\begin{equation}
\dot{q}^{i}=\partial _{p_{i}}H,\qquad \dot{p}_{i}=-\partial
_{q^{i}}H+F_{i}\,.  \label{otvHam}
\end{equation}%
In the framework of the open Hamiltonian system (\ref{otvHam})\
dynamics of any function $A(q^{i},p_{i},t)$ is defined by the
{open }(dissipative and/or forced){\ evolution equation}:
\begin{equation}
\partial _{t}A=-[A,H]+F_{i}[A,q^{i}]\,,\,\qquad (\,[A,q^{i}]=-\partial
_{p_{i}}A)\,.  \label{otv}
\end{equation}

In particular, if $A=\rho(q^{i},p_{i},t)$ represents the density function of
ensemble phase--points then its dynamics is given by the {open }%
(dissipative and/or forced) {Liouville equation}
\cite{GeoDyn,GaneshADG}:
\begin{equation}
\partial_{t}\rho=-[\rho,H]+F_{i}[\rho,q^{i}]\,.  \label{otvLio}
\end{equation}

{Equation} (\ref{otvLio}) {represents the open classical model of
our microscopic NN-dynamics.}

\subsection{Continuous neural network dynamics}

The generalized NN--dynamics, including two special cases of
{graded response neurons} (GRN) and {coupled neural oscillators}
(CNO), can be presented in the form of a {Langevin stochastic
equation} \cite{NeuFuz,ComNL}
\begin{equation}
\dot{\sigma}_{i}=f_{i}+\eta _{i}(t),  \label{langev}
\end{equation}%
where $\sigma _{i}=\sigma _{i}(t)$ are the continual neuronal variables of $i
$th neurons (representing either membrane action potentials in case of GRN,
or oscillator phases in case of CNO); $J_{ij}$ are individual synaptic
weights; $f_{i}=f_{i}(\sigma _{i},J_{ij})$ are the deterministic forces
(given, in GRN-case, by
\begin{equation*}
f_{i}=\sum_{j}J_{ij}\tanh [\gamma \sigma _{j}]-\sigma _{i}+\theta
_{i},\qquad \text{with \ }\gamma >0
\end{equation*}
and with the $\theta _{i}$ representing injected currents, and in CNO--case,
by
\begin{equation*}
f_{i}=\sum_{j}J_{ij}\sin (\sigma _{j}-\sigma _{i})+\omega _{i},
\end{equation*}
with $\omega _{i}$ representing the natural frequencies of the individual
oscillators); the noise variables are given as
\begin{equation*}
\eta _{i}(t)=\lim_{\Delta \rightarrow 0}\zeta _{i}(t)\sqrt{2T/\Delta },
\end{equation*}
where $\zeta _{i}(t)$ denote uncorrelated Gaussian distributed random forces
and the parameter $T$ controls the amount of noise in the system, ranging
from $T=0$ (deterministic dynamics) to $T=\infty $ (completely random
dynamics).

More convenient description of the neural random process
(\ref{langev}) is provided by the Fokker-Planck equation
describing the time evolution of the probability density $P(\sigma
_{i})$ \cite{StrAtr,LifeSpace,ComNL}
\begin{equation}
\partial _{t}P(\sigma _{i})=-\sum_{i}\partial _{\sigma _{i}}[f_{i}P(\sigma
_{i})]+T\sum_{i}\partial _{\sigma _{i}^{2}}P(\sigma _{i}).  \label{fp}
\end{equation}

Now, in the case of deterministic dynamics $T=0$, equation (\ref{fp}) can be
easily put into the form of the conservative Liouville equation (\ref{LioCl}%
), by making the substitutions:
\begin{equation*}
P(\sigma _{i})\rightarrow \rho ,f_{i}=\dot{\sigma}_{i},\qquad \text{and}%
\qquad \lbrack \rho ,H]=\mathrm{div}(\rho \,\dot{\sigma}_{i})\equiv
\sum_{i}\partial _{\sigma _{i}}\left( \rho \,\dot{\sigma}_{i}\right) ,
\end{equation*}
where $H=H(\sigma _{i},J_{ij})$. Further, we can formally identify the
stochastic forces, i.e., the second-order noise-term $T\sum_{i}\partial
_{\sigma _{i}^{2}}\rho $ with $F^{i}[\rho ,\sigma _{i}]\,$, to get the open
Liouville equation (\ref{otvLio}).

Therefore, on the NN--level deterministic dynamics corresponds to the
conservative system (\ref{LioCl}). Inclusion of stochastic forces
corresponds to the system opening (\ref{otvLio}), implying the {%
macroscopic arrow of time}.

\subsection{Open quantum system}

By formal quantization of equation (\ref{otvLio}), we obtain the quantum open Liouville
equation \cite{ComDyn,QuLeap}
\begin{equation}
\partial _{t}\hat{\rho}=i\{\hat{\rho},\hat{H}\}-i\hat{F}_{i}\{\hat{\rho},%
\hat{q}^{i}\}\,,  \label{otvQuLio}
\end{equation}
where $\hat{F}_{i}=\hat{F}_{i}(\hat{q}^{i},\hat{p}_{i},t)$ represents the
covariant quantum operator of external friction forces in the Hilbert state
space $\mathcal{H}=\mathcal{H}_{\hat{q}^{i}}\otimes \mathcal{H}_{\hat{p}%
_{i}} $.

{Equation} (\ref{otvQuLio}) {represents the open quantum-friction
model of our microscopic MT--dynamics.}

\subsection{Non--critical stringy MT--dynamics}

In EMN--language of non-critical (SUSY) bosonic strings, our
MT--dynamics equation (\ref{otvQuLio}) reads
\cite{ComDyn,GaneshADG,QuLeap}
\begin{equation}
\partial _{t}\hat{\rho}=i\{\hat{\rho},\hat{H}\}-i\hat{g}_{ij}\{\hat{\rho},%
\hat{q}^{i}\}\hat{\dot{q}}^{j}\,,  \label{emn}
\end{equation}
where the target-space density matrix $\hat{\rho}(\hat{q}^{i},\hat{p}_{i})$
is viewed as a function of coordinates $\hat{q}^{i}$ that parameterize the
couplings of the generalized $\sigma $-models on the bosonic string
world-sheet, and their conjugate momenta $\hat{p}_{i}$, while $\hat{g}_{ij}=%
\hat{g}_{ij}(\hat{q}^{i})$ is the quantum operator of the
{positive definite metric} in the space of couplings. Therefore,
the covariant quantum
operator of external friction forces is in EMN--formulation given as $\hat{F}%
_{i}(\hat{q}^{i},\hat{\dot{q}}^{i})=\hat{g}_{ij}\,\hat{\dot{q}}^{j}$.

Equation (\ref{emn}) establishes the conditions under which a large--scale
coherent state appearing in the MT-network, which can be considered
responsible for loss--free energy transfer along the tubulins.

The system-independent properties of equation (\ref{emn}), are:\newline
(i) Conservation of probability $P$
\begin{equation}
\partial _{t}P=\partial _{t}[\mathrm{Tr}(\hat{\rho})]=0.  \label{24}
\end{equation}%
(ii) Conservation of energy $E$, on the average
\begin{equation}
\partial _{t}\left\langle \left\langle E\right\rangle \right\rangle \equiv
\partial _{t}[\mathrm{Tr}(\hat{\rho}E)]=0.  \label{25}
\end{equation}%
(iii) Monotonic increase in entropy
\begin{equation}
\partial _{t}S=\partial _{t}[-\mathrm{Tr}(\hat{\rho}\ln \hat{\rho})]=(\hat{%
\dot{q}}^{i}\hat{g}_{ij}\hat{\dot{q}}^{j})S\geq 0,  \label{26}
\end{equation}%
due to the positive definiteness of the metric $\hat{g}_{ij}$, and
thus automatically and naturally implying a {microscopic arrow of
time} \cite{L}.

\subsection{Equivalence of Neurodynamic forms}

Both the macroscopic NN--equation (\ref{otvLio}) and the
microscopic MT--equation (\ref{otvQuLio}) have the same open
Liouville form, which implies the arrow of time \cite{GaneshADG,QuLeap}.
These demonstrates the existence of a formal neuro-biological
space-time self--similarity.

\section{Conclusion}

We have described neuro-- and psycho--dynamics of both NN and MT
ensembles, belonging to completely different biophysical
space-time scales, brain's neural networks and brain's
microtubules, by the unique form of the open Liouville equation,
which implies the arrow of time. In this way the existence of the
formal {neuro-biological space-time self-similarity} has been
proved.

\end{document}